\documentclass[11pt]{article}

\usepackage[utf8]{inputenc}
\usepackage[T1]{fontenc}
\usepackage{geometry}
\usepackage{amsmath,amssymb,bm,physics}
\usepackage{graphicx}
\usepackage[super,sort&compress]{natbib}
\usepackage[colorlinks=true,citecolor=blue,linkcolor=blue,urlcolor=blue]{hyperref}
\usepackage{authblk}
\usepackage{setspace}
\title{\bfseries Nonlocal transfer of quantized toroidal magnetic flux}

\author[1]{Adel Ali}
\author[1]{Alexey Belyanin}
\affil[1]{Department of Physics and Astronomy, Texas A\&M University, College Station, Texas 77843, USA}

\date{}

\begin{document}

\maketitle

\begin{abstract}
We propose a nonlocal flux-transfer experiment in which a quantized magnetic-field excitation confined within one toroidal superconducting structure is coherently transferred to a spatially separated toroid without magnetic-field occupation of the intervening region. The transfer arises from quantized Aharonov-Bohm-type vector potential coupling mediated by a superconducting loop threading the toroids, which, however, remains in the ground state, acting only through a global fluxoid constraint. A direct experimental signature would be the observation of correlated, time-resolved flux exchange between remote toroids in a SQUID readout. We analyze an apparent signaling paradox related to this interaction as a probe of the broader question of whether spatiotemporal quantum coherence is fundamentally bounded. The proposed setup can provide an experimental testbed for addressing foundational questions such as the existence of an objective collapse of a wavefunction or the fundamental limits of macroscopic quantum coherence which is relevant to large scale quantum computers.    
\end{abstract}

\section*{Introduction}

Progress in foundations of the quantum theory is often limited by the scarcity of near-term, technologically feasible experiments that sharply distinguish competing interpretations. Superconducting circuits, developed primarily for quantum information processing, now provide an exceptional platform for realizing and controlling quantum phenomena at macroscopic scale. This may help designing controlled tests of foundational ideas, including quantum nonlocality, in regimes where the relevant degrees of freedom, such as flux, charge, and phase, can be engineered, measured, and coherently manipulated.

Quantum mechanics assigns direct physical significance to electromagnetic potentials, most strikingly through the Aharonov--Bohm (AB) effect, in which charged matter acquires a measurable phase in regions where the local magnetic field vanishes. Experiments with toroidal magnetic structures and superconducting shielding established this point in a particularly compelling form by demonstrating phase shifts even when the magnetic field is confined and excluded from the electron trajectories \cite{Tonomura1986,Osakabe1986}. These results motivate a broader question that is both foundational and experimentally relevant in superconducting quantum devices: to what extent can one engineer coherent transfer of quantized flux excitations through couplings mediated by vector potentials, while keeping the real magnetic field confined to isolated regions of space?

In parallel, superconducting circuits have provided a mature platform for coherent quantum control, long-range interactions, and time-resolved readout. In circuit QED, spatially separated artificial atoms can exchange quantum information through a mediating mode, including in regimes where the transfer is governed by virtual rather than real excitations of the bus \cite{Majer2007,Blais2021}. Flux-sensitive readout using SQUID-based architectures is likewise well established and enables direct monitoring of superconducting flux degrees of freedom with high temporal resolution \cite{Lupascu2004}. These developments suggest that superconducting circuitry is an ideal setting in which to revisit AB-type nonlocal couplings in a fully quantum, controllable, and metrologically precise regime.

Here we propose a nonlocal flux-transfer experiment in which a localized magnetic-field excitation, trapped within one toroidal superconducting structure, is coherently transferred to a spatially separated toroid without magnetic-field occupation of the intervening region. The transfer is mediated by the AB-type vector-potential coupling of the quantized toroidal-flux degrees of freedom to a superconducting loop threading the toroids, while the physical magnetic field remains confined to the toroidal regions throughout the dynamics; see Fig.~1. In the regime of interest, the superconducting loop acts through a global fluxoid constraint, providing an effective interaction channel that correlates remote toroids without requiring propagating magnetic-field support in the space between them. Toroidal flux architecture has been proposed before for realizing exceptionally low-noise superconducting qubits \cite{Zagoskin2015} and field-free nonlocal coupling schemes involving toroidal flux qubits \cite{AliBelyanin2024}. The present proposal targets a distinct observable: time-resolved exchange of a localized flux excitation between remote toroidal traps.

The central experimental signature is a correlated oscillatory flux signal measured by spatially separated SQUID detectors attached to the two toroids, accompanied by a null test for magnetic-field leakage in the intervening region. Operationally, the protocol compares local flux readout at each toroid with direct field probes placed between the toroids, thereby distinguishing coherent transfer of the confined excitation from conventional stray-field crosstalk. Because the effect is mediated by a gauge-coupled superconducting network, the experiment also provides a route to probe the interplay between gauge potentials, confinement of electromagnetic fields, and coherent dynamics in mesoscopic quantum circuits.

Finally, we address an apparent signaling paradox suggested by the nonlocal character of the effective coupling. Although the transfer process can produce remote, time-resolved correlations that may naively resemble instantaneous action, any consistent description must remain compatible with operational no-signaling and relativistic causality \cite{Simon2001,Brunner2014}. We outline a consistent resolution and identify the parameter regime, timing constraints, and measurement protocol required for a decisive experimental test.

A central question is whether quantum coherence can be sustained as a genuinely spatiotemporal resource, simultaneously extended over macroscopic distances and preserved for operationally long times, or whether fundamental constraints ultimately bound such nonlocal coherence independently of engineering details. This question is particularly timely in the context of emerging quantum technologies, and especially for large-scale quantum computing architectures that aim to use long-lived, spatially extended coherent quantum states as a computational resource. In such platforms, useful quantum information processing requires the preparation, control, and preservation of coherence across many degrees of freedom and over operationally relevant timescales. Understanding whether there exist fundamental limits on this spatiotemporal coherence, beyond device-specific sources of noise and decoherence, is therefore a central issue with direct implications for scalability and fault tolerance. Related questions have been addressed in the context of macroscopic quantum coherence and the quantum-to-classical transition \cite{Leggett1980,JoosZeh1985,Zurek2003,Schlosshauer2007,ArndtHornberger2014,NimmrichterHornberger2013}, phenomenological collapse models \cite{GRW1986,Pearle1989,GhirardiPearleRimini1990,BassiGhirardi2003,BassiRMP2013}, gravity-induced decoherence or reduction \cite{Diosi1987,Penrose1996,Pikovski2015,Blencowe2013,Donadi2021}, and operational causality bounds on macroscopic superpositions \cite{Mari2016,Belenchia2018}.

The structure of the paper is as follows. First, we describe an experiment in which the energy of the magnetic field trapped inside a finite volume of space is transferred to another trapped finite volume of space without traversing the space in between. Second, we use a charged-ring and fluxoid-loop model to expose the causal assumptions behind the reduced Hamiltonian and to formulate a collapse criterion for long-range spatiotemporal coherence.


\begin{figure}
    \centering
    \includegraphics[width=0.9\linewidth]{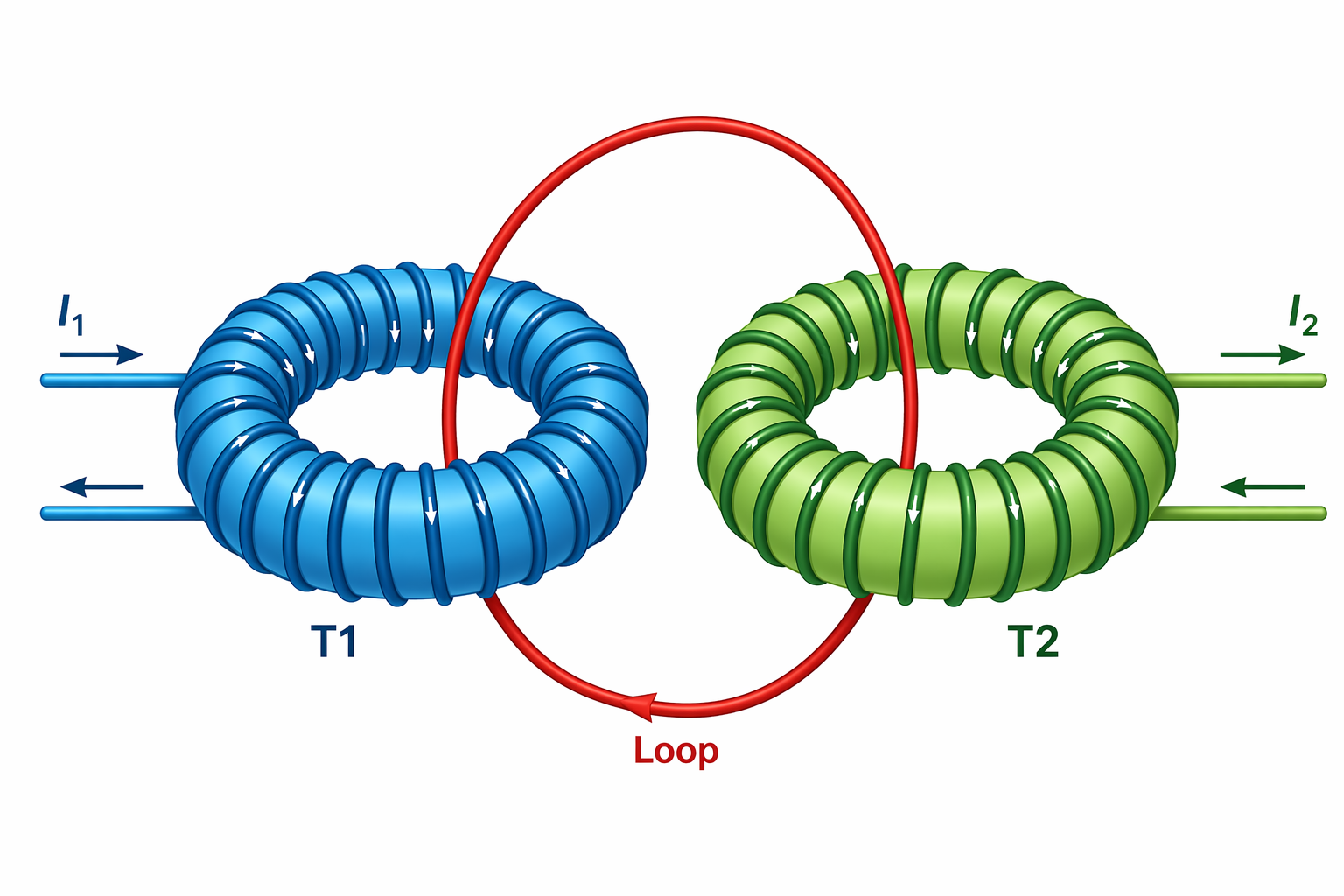}
    \caption{
    Schematic of two toroidal superconducting flux elements interlinked by a common superconducting loop. The magnetic field of each toroid is confined to its core, while the linking loop is sensitive to the collective fluxoid in Eq.~(\ref{eq:linked_flux}). The nonlocal transfer of the toroidal flux is mediated by a global fluxoid constraint imposed by the superconducting loop. Local SQUID detectors measure the toroidal fluxes, and field probes placed between the toroids perform a null test for the magnetic-field leakage. 
    }
    \label{fig:placeholder}
\end{figure}


\section*{Results}

\subsection*{Fluxoid mediated coupling of toroidal flux modes}

A central question, with immediate applications in superconducting quantum circuits, is whether two spatially separated flux degrees of freedom can exchange quantum energy through a channel containing no appreciable magnetic field in the intermediate region. Toroidal superconductors provide a natural setting for this question. Their magnetic flux can be confined inside the toroidal cores, while a superconducting loop threading the two holes remains sensitive to the associated gauge holonomy. The coupling is therefore not set by a local stray magnetic field, but by the fluxoid constraint of the linking loop. This geometry realizes a minimal circuit-QED problem in which two toroidal flux modes are coupled by a global superconducting phase constraint. Equivalently, a change of flux in one toroidal element appears as a force on the other through a shared topological boundary condition.

We consider two toroidal flux modes, labeled \(j=2,3\), linked by a common superconducting loop with inductance \(L\); see Fig.~1. The magnetic fields of the toroids are assumed to be confined inside their cores, but the linking loop encloses the corresponding fluxes. The total linked flux is
\begin{equation}
    \Phi_{\rm link}
    =
    \Phi_{\rm b}
    +
    s_2\hat\phi_2
    +
    s_3\hat\phi_3 ,
    \label{eq:linked_flux}
\end{equation}
where \(\Phi_{\rm b}\) is a classical bias flux and \(s_j=\pm1\) are the oriented linking numbers. The superconducting phase around the loop obeys
\begin{equation}
    \oint_{\mathcal C}\nabla\theta\cdot d{\bf l}=2\pi n ,
    \qquad n\in\mathbb Z ,
     \nonumber 
\end{equation}
so that, in a fixed fluxoid sector \(n\), the persistent current is
\begin{equation}
    I_n
    =
    \frac{
    n\Phi_0-\Phi_{\rm link}
    }{L},
    \quad
    \Phi_0=\frac{h}{2e}.
     \nonumber 
\end{equation}
The inductive energy of the linking loop is therefore
\begin{equation}
    H_{\rm link}^{(n)}
    =
    \frac{
    \left[
    n\Phi_0-\Phi_{\rm b}
    -s_2\hat\phi_2
    -s_3\hat\phi_3
    \right]^2
    }{2L}.
\end{equation}
Defining
\begin{equation}
    \delta\Phi_n\equiv n\Phi_0-\Phi_{\rm b},
     \nonumber 
\end{equation}
we may write
\begin{equation}
    H_{\rm link}^{(n)}
    =
    \frac{
    \left[
    \delta\Phi_n
    -
    \hat X
    \right]^2
    }{2L},
    \quad
    \hat X=s_2\hat\phi_2+s_3\hat\phi_3 .
     \nonumber 
\end{equation}
Thus the linking loop couples only to the collective topological flux \(\hat X\), rather than to the local magnetic-field profile of either toroid.

The total Hamiltonian is
\begin{equation}
    H^{(n)}
    =
    H_{\rm T}^{(2)}
    +
    H_{\rm T}^{(3)}
    +
    \frac{
    \left[
    \delta\Phi_n
    -s_2\hat\phi_2
    -s_3\hat\phi_3
    \right]^2
    }{2L},
    \label{eq:general_fluxoid_hamiltonian}
\end{equation}
where \(H_{\rm T}^{(j)}\) denotes the internal Hamiltonian of the \(j\)th toroidal flux element. Expanding the last term gives
\begin{align}
    H^{(n)}
    =
    H_{\rm T}^{(2)}
    +
    H_{\rm T}^{(3)}
    -
    \frac{\delta\Phi_n}{L}
    \left(
    s_2\hat\phi_2+s_3\hat\phi_3
    \right)
    +
    \frac{\hat\phi_2^2+\hat\phi_3^2}{2L}
    +
    \frac{s_2s_3}{L}\hat\phi_2\hat\phi_3
    +{\rm const}.
     \nonumber 
\end{align}
The bias term acts as a tunable longitudinal force on each toroidal flux coordinate, while the last term is the desired nonlocal fluxoid-mediated interaction. Its sign is fixed by the relative linking orientation, and its magnitude is controlled by the inductance \(L\) of the common loop.

\subsection*{Harmonic toroidal modes}

We first specialize Eq.~\eqref{eq:general_fluxoid_hamiltonian} to the case in which each toroidal flux element is a harmonic oscillator,
\begin{equation}
    H_{\rm T}^{(j)}
    =
    \frac{\hat q_j^2}{2C_j}
    +
    \frac{\hat\phi_j^2}{2L_j},
    \qquad
    \omega_j=\frac{1}{\sqrt{L_jC_j}} .
     \nonumber 
\end{equation}
The full Hamiltonian is then
\begin{equation}
    H_{\rm osc}^{(n)}
    =
    \sum_{j=2,3}
    \left(
    \frac{\hat q_j^2}{2C_j}
    +
    \frac{\hat\phi_j^2}{2L_j}
    \right)
    +
    \frac{
    \left[
    \delta\Phi_n
    -s_2\hat\phi_2
    -s_3\hat\phi_3
    \right]^2
    }{2L}.
     \nonumber 
\end{equation}
Expanding, one obtains
\begin{align}
    H_{\rm osc}^{(n)}
    =
    \sum_{j=2,3}
    \frac{\hat q_j^2}{2C_j}
    +
    \frac{1}{2}
    \sum_{j=2,3}
    K_j\hat\phi_j^2
    +
    J\hat\phi_2\hat\phi_3 -
    \sum_{j=2,3}F_j\hat\phi_j
    +
    {\rm const},
     \nonumber 
\end{align}
with
\begin{equation}
    K_j=\frac{1}{L_j}+\frac{1}{L},
    \qquad
    J=\frac{s_2s_3}{L},
    \qquad
    F_j=\frac{s_j\delta\Phi_n}{L}.
     \nonumber 
\end{equation}
Thus the fluxoid constraint converts two independent harmonic flux oscillators into a pair of linearly coupled normal modes. The coupling strength is fixed by the inductance of the linking loop and by the relative orientation of the two linkings.

The Heisenberg equations of motion have the same form as the classical equations for this quadratic Hamiltonian,
\begin{align}
    C_2\ddot{\hat\phi}_2
    +
    K_2\hat\phi_2
    +
    J\hat\phi_3
    &=
    F_2,
    \label{eq:eom_phi2}
    \\
    C_3\ddot{\hat\phi}_3
    +
    K_3\hat\phi_3
    +
    J\hat\phi_2
    &=
    F_3 .
    \label{eq:eom_phi3}
\end{align}
Equations~\eqref{eq:eom_phi2} and \eqref{eq:eom_phi3} display the essential physics: any time-dependent change of \(\phi_2(t)\) appears as a driving force on \(\phi_3(t)\),
\begin{equation}
    C_3\ddot\phi_3+K_3\phi_3
    =
    F_3
    -
    \frac{s_2s_3}{L}\phi_2(t).
     \nonumber 
\end{equation}
Similarly, \(\phi_3(t)\) appears in the equation of motion for \(\phi_2(t)\). The coupling is reciprocal and originates from the collective combination \(s_2\phi_2+s_3\phi_3\) in the total Hamiltonian.

In frequency space, the response of oscillator \(3\) to a given Fourier amplitude \(\delta\phi_2(\Omega)\) is
\begin{equation}
    \delta\phi_3(\Omega)
    =
    -
    \frac{J}{K_3-C_3\Omega^2}
    \delta\phi_2(\Omega),
    \label{eq:transfer_function}
\end{equation}
provided backaction on oscillator \(2\) is neglected. Thus every Fourier component of the flux motion in one toroid produces a corresponding response in the other, with susceptibility set by the dressed oscillator pole of the second toroid. 

After shifting to the equilibrium configuration set by \(\delta\Phi_n\), the fluctuation Hamiltonian is
\begin{equation}
    H_{\rm fluc}
    =
    \sum_{j=2,3}
    \left[
    \frac{\delta q_j^2}{2C_j}
    +
    \frac{K_j}{2}\delta\phi_j^2
    \right]
    +
    J\delta\phi_2\delta\phi_3 .
     \nonumber 
\end{equation}
The normal-mode frequencies follow from
\begin{equation}
    \left(K_2-C_2\Omega^2\right)
    \left(K_3-C_3\Omega^2\right)
    -J^2=0 .
     \label{eq:normal_mode_equation}
\end{equation}
For identical toroidal oscillators, \(C_2=C_3=C_{\rm T}\) and \(L_2=L_3=L_{\rm T}\), the two collective normal modes are 
\begin{equation}
    \Omega_\pm^2
    =
    \frac{1}{C_{\rm T}}
    \left[
    \frac{1}{L_{\rm T}}
    +
    \frac{1}{L}
    \pm
    \frac{s_2s_3}{L}
    \right].
    \label{eq:normal_modes_identical}
\end{equation}
The relative orientation \(s_2s_3\) determines which collective combination is shifted upward or downward. The resulting mode splitting is a direct observable signature of the fluxoid-mediated coupling.

\subsection*{Two level toroidal flux qubits}

We now project the same fluxoid Hamiltonian onto the two lowest states of each toroidal flux element. This typically implies that a nonlinearity is introduced (e.g., Josephson junctions) to make energy levels nonequidistant. In the persistent-current basis,
\begin{equation}
    \hat\phi_j
    \simeq
    \phi_j^\ast \sigma_j^z ,
    \nonumber 
\end{equation}
where \(\phi_j^\ast\) is the flux matrix element of the \(j\)th toroidal qubit. Including the intrinsic tunneling splitting \(\Delta_j\), the effective two-qubit Hamiltonian becomes
\begin{equation}
    H_{\rm q}^{(n)}
    =
    -\sum_{j=2,3}
    \frac{\Delta_j}{2}\sigma_j^x
    -
    \sum_{j=2,3}
    \frac{s_j\delta\Phi_n\phi_j^\ast}{L}
    \sigma_j^z
    +
    J_{zz}\sigma_2^z\sigma_3^z ,
    \label{eq:two_qubit_hamiltonian}
\end{equation}
with
\begin{equation}
    J_{zz}
    =
    \frac{s_2s_3\phi_2^\ast\phi_3^\ast}{L}.
      \nonumber 
\end{equation}
The terms \(\hat\phi_j^2/(2L)\) generated by the linking loop become constants under this ideal two-level projection and have therefore been absorbed into the energy offset. Equation~\eqref{eq:two_qubit_hamiltonian} shows that the common superconducting loop converts the linked toroidal fluxes into an Ising-type interaction. At the fluxoid-symmetry point \(\delta\Phi_n=0\), the longitudinal bias vanishes while the nonlocal interaction remains. Away from this point, the classical bias \(\Phi_{\rm b}\) controls the qubit detunings and can be used to bring the two qubits into or out of resonance.

The statement that a change of one toroidal flux ``appears'' in the other means that the two variables are coupled by the same global superconducting fluxoid constraint. It does not imply instantaneous propagation of a local electromagnetic disturbance. The transfer time scale depends on the transverse coupling parameter \(J_{zz}\).

The essential physical consequence is that two flux modes with locally confined magnetic fields can nevertheless interact through a shared superconducting fluxoid constraint. The coupling is controlled by topology and circuit inductance rather than by ordinary mutual inductance between stray magnetic fields. Removing the linking loop, or changing the topology so that the toroids are not linked by the common superconducting path, eliminates the effect. Conversely, preserving the linking while deforming the toroids or the loop changes microscopic details but leaves the leading fluxoid coupling invariant.

\subsection*{No signaling assumptions and apparent Hamiltonian nonlocality}

Bell tests establish that quantum mechanics permits correlations that are incompatible with local hidden-variable models. At the same time, the no-communication principle guarantees that spatially separated parties sharing an arbitrary joint state, including maximally entangled states, cannot transmit controllable information by local operations on their respective subsystems.

A useful perspective from quantum information is that several cornerstone constraints of quantum theory can be viewed as consistency requirements of no-signaling. A standard example is the no-cloning theorem: if perfect cloning were possible, then Alice and Bob could convert a choice of measurement basis into a distinguishable ensemble on Bob's side, enabling superluminal signaling. For example, let Alice and Bob share a Bell pair \( |\Phi^+\rangle=(|00\rangle+|11\rangle)/\sqrt{2}\), which is equivalently \( |\Phi^+\rangle=(|++\rangle+|--\rangle)/\sqrt{2}\), where \(|\pm\rangle\) are eigenstates along some Bloch-sphere axis. If Bob could clone his reduced state, then by measuring many clones he could discriminate which decomposition, computational or rotated basis, Alice has prepared by her local choice, thereby extracting information from spacelike separation. This contradiction enforces the impossibility of perfect cloning under unitary quantum evolution \cite{cusalcomplexity,cloningCTC}. In this sense, attempts to violate no-signaling tend to collide with other basic postulates, here unitarity and linearity.

Related arguments have also been used to motivate or constrain measurement postulates. In particular, if one modifies the standard assignment of probabilities to measurement outcomes, or assumes that reduced density matrices fail to encode all operational statistics, one generically opens a channel for signaling; conversely, imposing no-signaling can strongly restrict admissible probability rules, providing a route toward the Born rule \cite{born}. These examples suggest a broader theme: systems and protocols that appear to evade no-signaling typically do so only by implicitly abandoning another foundational quantum postulate.

The no-communication theorem is usually stated under an additional structural assumption: each party's admissible operations are generated by local interactions acting on their own degrees of freedom \cite{localinteraction}. We therefore ask what are the operational consequences of relaxing this locality assumption in systems where the effective interaction is intrinsically nonlocal, not in the Bell sense of entanglement-enabled correlations, but in the sense that one subsystem's Hamiltonian depends on a remote quantum degree of freedom?

Aharonov--Bohm physics provides a natural arena for such questions. In the magnetic AB effect, an electron acquires a phase determined by an enclosed flux even when it propagates through regions of vanishing magnetic field. This field-free sensitivity is distinct from Bell nonlocality and does not require entanglement. It has long motivated speculations about signaling, but careful analyses show that in the conventional semiclassical setting any rapid modulation of the magnetic flux is accompanied by an induced electric field whose contribution cancels the would-be superluminal effect \cite{abftl}. Prior discussions typically treat the flux source classically and the electron as a scattering degree of freedom \cite{scattering}.

Here we focus instead on a fully quantum AB-type setting in which both the matter and the enclosed flux are dynamical quantum degrees of freedom. In such a description, the AB coupling appears as an emergent interaction between subsystems that is topological in origin and mediated by a quantized vector potential in regions free of classical fields. This can be summarized by the following thought experiment. Consider two distant parties, Alice, system \(A\), and Bob, system \(B\), prepared in the ground state of a global Hamiltonian \(H\). Even if the ground state factorizes, \( |\Psi_0\rangle=|\psi_A\rangle\otimes|\psi_B\rangle\), a signaling channel becomes possible if Bob can perform a local operation that instantaneously changes the Hamiltonian governing Alice's subsequent evolution: \(H\to H'\), with \(H'\) acting differently on \(A\). This is conceptually trivial, but it isolates the precise assumption being challenged: admissible operations do not enact nonlocal changes of the generator of time evolution.


In the AB-flavored architecture discussed below, the emergent exchange coupling between the toroidal fluxes can grow with the number of particles on the ring. This implies that for a critical number of particles at fixed circumference, a flux quanta can be exchanged faster than the speed of light in vacuum. This apparent paradox does not contradict the standard no-communication theorem because it violates one of its premises: Alice's reduced dynamics is not generated by a strictly local Hamiltonian independent of Bob's subsystem. The central question is then not whether signaling is formally possible in this model, but rather which physical principle forbids such instantaneous Hamiltonian switching. 
The phase-only protocol discussed in the Methods is designed to separate the usual Faraday-field channel from a genuine flux-transfer event. A possibility, which we highlight as experimentally testable, is that macroscopic spatial superpositions, or long-range spatiotemporal coherence of the many-body wavefunction, are limited beyond a characteristic spacetime volume, effectively preventing the assumed coherent, global operation from being physically realizable. This is because the spatial coherence of the extended many-body wavefunction is necessary for mediating the coupling. 

Any outcome of testing this hypothesis is of foundational and practical interest. Foundationally, the question intersects long-standing debates on collapse, locality, and the physical meaning of the wave function, but now in a setting amenable to superconducting-circuit experiments. Practically, if there exist fundamental constraints on coherent quantum evolution over large spacetime volumes, they would impose hard limits on the scalable quantum volume achievable in quantum processors, independent of engineering noise sources, since sufficiently large coherent machines could otherwise implement operations that emulate nonlocal Hamiltonian control.

\subsection*{Bosonic ring and superconducting loop mediators}

For the purpose of this section we first replace the superconducting loop by
charged bosons on a ring. This makes the derivation conceptually transparent.
The superconducting-loop realization is obtained below as the corresponding
kinetic-inductance fluxoid problem and leads to the same effective structure.
We also treat the toroids as two-level flux qubits, similarly to the
two-level toroidal flux-qubit model above. This typically requires introducing
nonlinearity, for example Josephson junctions or phase slips.

We consider \(N\) identical charged bosons of mass \(M\) on a ring of radius
\(R\), with single-particle rotor scale
\begin{equation}
E_R=\frac{\hbar^2}{2MR^2}.
  \nonumber 
\end{equation}
The ring Hamiltonian is
\begin{equation}
\hat H_{\rm ring}
=
E_R\sum_{m\in\mathbb Z}
\bigl(m-\beta-\hat N_q\bigr)^2
\hat b_m^\dagger \hat b_m,
\label{eq:Hring}
\end{equation}
where \(\hat b_m^\dagger\) creates a boson in angular-momentum orbital
\(m\in\mathbb Z\),
\begin{equation}
\hat n^{(i)}=\frac{1-\sigma_z^{(i)}}{2},
\qquad
\hat N_q=\hat n^{(2)}+\hat n^{(3)},
  \nonumber 
\end{equation}
and \(\beta=\Phi_{\rm b}/\Phi_0\). We assume that each flux qubit contributes
one flux quantum, so that the reduced flux shift seen by the ring is an
integer. Equation~\eqref{eq:Hring} is the strict one-dimensional version of
the fluxoid-link Hamiltonian, with the inductive energy expressed as the
kinetic energy of charged particles on the ring.

Including the tunnelling of the two flux qubits, the full Hamiltonian is
\begin{equation}
\hat H=\hat H_{\rm ring}
+\Delta\bigl(\sigma_x^{(2)}+\sigma_x^{(3)}\bigr).
  \nonumber 
\end{equation}
Projecting onto the fixed ring state in which all bosons occupy \(m=0\),
\begin{equation}
|N_0\rangle=\frac{(\hat b_0^\dagger)^N}{\sqrt{N!}}|{\rm vac}\rangle,
  \nonumber 
\end{equation}
gives the reduced qubit Hamiltonian
\begin{equation}
\hat H_{N_0}
=
N E_R\bigl(\beta+\hat N_q\bigr)^2
+\Delta\bigl(\sigma_x^{(2)}+\sigma_x^{(3)}\bigr).
\label{eq:Hnaive}
\end{equation}
Expanding Eq.~\eqref{eq:Hnaive} gives 
\begin{equation}
\hat H_{N_0}
=
\varepsilon_0
+h_0\bigl(\sigma_z^{(2)}+\sigma_z^{(3)}\bigr)
+J_N\sigma_z^{(2)}\sigma_z^{(3)}
+\Delta\bigl(\sigma_x^{(2)}+\sigma_x^{(3)}\bigr),
  \nonumber 
\end{equation}
with
\begin{align}
J_N =\frac{N E_R}{2}, \quad
h_0=-N E_R(\beta+1), \quad 
\varepsilon_0 = N E_R\left[(\beta+1)^2+\frac12\right].
  \nonumber 
\end{align}
At the exchange point \(h_0=0\), namely \(\beta=-1\), we rotate to the qubit
energy basis,
\begin{equation}
\tau_z^{(i)}=\sigma_x^{(i)},
\qquad
\tau_x^{(i)}=\sigma_z^{(i)}.
  \nonumber 
\end{equation}
The projected Hamiltonian becomes
\begin{equation}
\hat H_{N_0}
=
\varepsilon_0
+\Delta\bigl(\tau_z^{(2)}+\tau_z^{(3)}\bigr)
+J_N\tau_x^{(2)}\tau_x^{(3)}.
\label{eq:HnaiveRot}
\end{equation}

The bosons do not have to dynamically scatter to higher \(m\) states to create
this coupling. Their static existence in the delocalized \(m=0\) state acts as
a mediator. Unlike conventional quantum-bus architectures, where the
qubit--qubit interaction is generated by virtual excitations of a dynamical
mode, the coupling here originates from a conserved many-body background. The
bosonic occupations commute with the full Hamiltonian, so a ring state prepared
in a fixed sector remains frozen throughout the evolution. The interaction is
therefore not mediated by propagation through an intermediate excitation, but
by the dependence of the ring energy on the collective qubit flux
configuration.

In the one-excitation subspace \(\{|eg\rangle,|ge\rangle\}\) of the
\(\tau_z\) basis,
\begin{equation}
\hat H_{\rm odd}
=
\varepsilon_0\mathbb I
+
J_N\left(
|eg\rangle\langle ge|
+
|ge\rangle\langle eg|
\right),
  \nonumber 
\end{equation}
so that
\begin{equation}
P_{2\to3}(t)
=
\sin^2\left(\frac{J_Nt}{\hbar}\right).
  \nonumber 
\end{equation}
The effective exchange rate is therefore enhanced by the particle number,
\(J_N\propto N\). This apparent scaling raises a potential causality paradox:
for sufficiently large \(N\), the projected flux-exchange process could become
spacelike for a fixed geometry and separation distance. This suggests that a
many-body correlated wavefunction cannot remain arbitrarily coherently
extended in spacetime. We therefore conjecture the existence of a fundamental
bound on the spatiotemporal coherence of the \(N\)-particle wavefunction,
beyond which the state undergoes spontaneous collapse upon reaching a critical
quantum volume.

The full swap time in the projected \(m=0\) sector is
\begin{equation}
t_{\rm swap}
=
\frac{\pi\hbar}{2J_N}
=
\frac{2\pi M R^2}{N\hbar}.
\nonumber 
\end{equation}
Requiring \(t_{\rm swap}\ge 2R/c\), where \(2R/c\) is the light-crossing time
across the ring diameter, gives
\begin{equation}
N_c=\frac{\pi M R c}{\hbar}.
\nonumber 
\end{equation}
The corresponding critical linear density is
\begin{equation}
n_c=\frac{N_c}{2\pi R}
=
\frac{M c}{2\hbar}.
\nonumber 
\end{equation}
Thus, within the projected model, the apparent superluminal threshold is set
by a radius-independent critical linear density that scales as an inverse
Compton wavelength.

The one-dimensional ring model isolates the conceptual tension in its simplest form; in an actual device a superconducting condensate is an example of these bosons, and the mediating condensate is a three-dimensional many-body state. The kinetic energy of the charged bosons is replaced by the kinetic inductive energy of the superconducting condensate. The same effective structure follows for a superconducting fluxoid-loop
mediator in the kinetic-inductance limit. Let the common superconducting loop have circumference \(s\), cross-sectional area \(A\), Cooper-pair mass \(M_\ast\simeq2m_e\), Cooper-pair charge \(q_\ast=2e\), and condensate density \(n_s\). Its kinetic inductance is 
\begin{equation} L_k = \frac{M_\ast s}{n_s q_\ast^2 A} = \frac{M_\ast s^2}{N_s q_\ast^2}, \qquad N_s=n_sAs . 
\nonumber 
\end{equation} 
Thus a larger participating condensate density reduces \(L_k\) and increases the fluxoid energy. In a fixed fluxoid sector \(\nu=\nu_0\), the superconducting-loop Hamiltonian has the same dependence on the collective qubit flux \(\hat N_q\) as the bosonic ring Hamiltonian, 
\begin{equation} \hat H_{\nu_0} = E_k \left( \nu_0-\beta-\hat N_q \right)^2 + \Delta\bigl(\sigma_x^{(2)}+\sigma_x^{(3)}\bigr), \label{eq:Hnu0} 
\end{equation} 
where \begin{equation} E_k = \frac{\Phi_0^2}{2L_k} = \frac{N_s h^2}{2M_\ast s^2}. 
  \nonumber 
\end{equation} 
Thus the fluxoid energy is the kinetic energy of the condensate required to maintain a single-valued superconducting phase in the presence of the enclosed Aharonov--Bohm flux. Comparing Eq.~\eqref{eq:Hnu0} with Eq.~\eqref{eq:Hnaive} shows that the superconducting-loop result is obtained from the bosonic-ring result by the replacement 
\begin{equation} N E_R\longrightarrow E_k . 
\nonumber \end{equation} 
Consequently, after expanding in Pauli matrices, one obtains the same longitudinal Ising form, \begin{equation} 
\hat H_{\nu_0} = \varepsilon_{\nu_0} + h_{\nu_0} \left( \sigma_z^{(2)}+\sigma_z^{(3)} \right) + J_k\sigma_z^{(2)}\sigma_z^{(3)} + \Delta \left( \sigma_x^{(2)}+\sigma_x^{(3)} \right), 
  \nonumber 
\end{equation} 
with 
\begin{align} J_k &= \frac{E_k}{2} = \frac{\Phi_0^2}{4L_k} = \frac{N_s h^2}{4M_\ast s^2},   \nonumber \\ 
h_{\nu_0} &= E_k(\nu_0-\beta-1), \quad  \varepsilon_{\nu_0} = E_k \left[ (\nu_0-\beta-1)^2+\frac{1}{2} \right]. 
  \nonumber 
  \end{align} 
  At the exchange point, the same rotation to the qubit energy basis gives 
  \begin{equation} 
  \hat H_{\nu_0}^{\rm ex} = \varepsilon_{\nu_0} + \Delta\bigl(\tau_z^{(2)}+\tau_z^{(3)}\bigr) + J_k\tau_x^{(2)}\tau_x^{(3)} . 
  \label{eq:HkineticRot} 
  \end{equation} Thus the one-excitation subspace has the same exchange dynamics as the bosonic-ring model, with \(J_N\) replaced by \(J_k\): \begin{equation} P_{2\rightarrow3}(t) = \sin^2\left(\frac{J_kt}{\hbar}\right). \end{equation} The corresponding full swap time is \begin{equation} t_{\rm swap} = \frac{\pi\hbar}{2J_k} = \frac{2\pi\hbar L_k}{\Phi_0^2} = \frac{L_k}{R_Q} = \frac{M_\ast s^2}{N_s h}, \label{eq:tswapKineticRQ} \end{equation} where \begin{equation} 
  R_Q = \frac{\Phi_0^2}{2\pi\hbar} = \frac{h}{(2e)^2}. 
  \nonumber 
  \end{equation} 

To remain self-consistent, the exchange time predicted by this Hamiltonian should not be shorter than the causal light-crossing time across the device. Taking \(L_{\rm coh}\) as the relevant spatial separation or coherence length, this requires 
\begin{equation} t_{\rm swap} \gtrsim \frac{L_{\rm coh}}{c}. 
  \nonumber \end{equation} 
  Using Eq.~\eqref{eq:tswapKineticRQ}, the participating Cooper-pair number must therefore satisfy 
  \begin{equation} N_s \lesssim N_s^{\rm crit} = \frac{M_\ast c}{h} \frac{s^2}{L_{\rm coh}} . 
    \nonumber 
  \end{equation} Equivalently, for \(n_{1D}=N_s/s\), \begin{equation} 
  n_{1D} \lesssim n_{1D}^{\rm crit} = \frac{M_\ast c}{h} \frac{s}{L_{\rm coh}} . 
    \nonumber 
  \end{equation} 
  For the natural choice \(L_{\rm coh}=s\), 
  \begin{equation} n_{1D}^{\rm crit} = \frac{M_\ast c}{h} \simeq 8.2\times10^{11}\ {\rm m}^{-1} \simeq 8.2\times10^5\ {\rm pairs}/\mu{\rm m}. 
  \nonumber 
  \end{equation}
  In terms of the three-dimensional superfluid density, \begin{equation} n_s 
  \lesssim n_s^{\rm crit} = \frac{1}{A} \frac{M_\ast c}{h} \frac{s}{L_{\rm coh}} . 
    \nonumber 
  \end{equation} 
  Thus, for a purely kinetic-inductance loop, the requirement that the projected exchange dynamics not outrun the light-crossing time becomes an upper bound on the participating superfluid density. For \(L_{\rm coh}=s\), \begin{equation} n_s \lesssim \frac{8.2\times10^{11}}{A}\ {\rm m}^{-3}. \end{equation} A wire with \(A=10^{-15}\ {\rm m}^2\) therefore has 
  \begin{equation} n_s^{\rm crit} \sim 8\times10^{26}\ {\rm m}^{-3}.   \nonumber 
  \end{equation} 
  Larger cross sections require proportionally smaller effective superfluid density in order to remain within the kinetic-inductance causal bound.

\subsection*{Causal collapse criterion}

The projected exchange dynamics implies a characteristic timescale that decreases linearly with the boson number. Interpreting this trend literally would eventually drive the nominal coherent dynamics below the light-crossing time, signaling a breakdown of the model. Motivated by this observation, we conjecture a causal collapse criterion. Let \(L_{\rm coh}\) denote the coherence length of the delocalized wavefunction and \(n=N/L_{\rm coh}\) its linear density. We postulate that the maximum duration over which the many-body state can remain coherent along the circumference of the ring is
\begin{equation}
\tau_{\max}(n,L_{\rm coh})
=
\frac{L_{\rm coh}}{c}\frac{n_c}{n}.
\label{eq:tau_max_causal}
\end{equation}
Equivalently, coherent unitary evolution is assumed to be restricted to the regime
\begin{equation}
\chi_{\rm caus}
\equiv
\frac{n}{n_c}\frac{c\,\tau_{\rm coh}}{L_{\rm coh}}
\lesssim 1,
\label{eq:chi_caus}
\end{equation}
where \(\tau_{\rm coh}\) is the coherence time of the delocalized many-body wavefunction. Once \(\chi_{\rm caus}\gtrsim 1\), the evolution is assumed not to be unitary and the state is assumed to undergo spontaneous collapse or be subjected to fundamental decoherence. In other words, non-unitarity or collapse is triggered when the spacetime footprint required to sustain coherent many-body evolution exceeds the causal coherence volume permitted by the density of the state. At the critical density \(n=n_c\), Eq.~\eqref{eq:tau_max_causal} reduces to the light-crossing bound \(\tau_{\max}=L_{\rm coh}/c\). For the ring geometry considered above, one has \(L_{\rm coh}=2R\), so that the causal bound becomes \(\tau_{\max}=(2R/c)(n_c/n)\).

\subsection*{Loss of long range spatial coherence on the ring}

The apparent superluminal scaling obtained from the projected exchange Hamiltonian suggests that the effective fully coherent description of the bosonic ring cannot remain valid arbitrarily far into the large-\(N\) regime. As a minimal resolution, we postulate that what breaks down is not the existence of the bosons themselves, but their global spatial coherence around the ring. In this picture, above a critical linear density the bosonic field can no longer sustain phase coherence over the full circumference for long duration of time, and the off-diagonal long-range order responsible for coherent mediation is progressively suppressed.

To express this idea in the simplest form, we introduce the bosonic field operator on the ring,
\begin{equation}
\hat\psi(\theta)
=
\frac{1}{\sqrt{2\pi}}
\sum_{m\in\mathbb Z}
e^{i m\theta}\,
\hat b_m,
\qquad
\theta\in[0,2\pi),
\nonumber 
\end{equation}
and the corresponding local density operator
\begin{equation}
\hat n(\theta)=\hat\psi^\dagger(\theta)\hat\psi(\theta).
\nonumber 
\end{equation}
The spatial coherence of the bosons is characterized by the one-body density matrix
\begin{equation}
G^{(1)}(\theta,\theta';t)
=
\langle
\hat\psi^\dagger(\theta,t)\hat\psi(\theta',t)
\rangle .
\label{eq:G1def}
\end{equation}
A globally coherent ring state exhibits sizable off-diagonal correlations for angular separations of order \(\abs{\theta-\theta'}\sim \pi\), whereas loss of long-range coherence is signaled by the decay of \(G^{(1)}(\theta,\theta';t)\) for \(\theta\neq\theta'\).

We model this effect by supplementing the unitary dynamics with a density-localizing stochastic process. At the ensemble level this is described by the master equation
\begin{align}
\partial_t \rho
& =
-\frac{i}{\hbar}[\hat H,\rho] -\frac{\Gamma(n)}{2}
\int_{0}^{2\pi} d\theta \int_{0}^{2\pi} d\theta'\,
K_\ell(\theta-\theta')
\bigl[\hat n(\theta),[\hat n(\theta'),\rho]\bigr],
\label{master_ring_collapse}
\end{align}
where \(K_\ell(\theta-\theta')\) is a positive translationally invariant kernel on the ring, \(\ell\) is a coherence length, and
\begin{equation}
n=\frac{N}{2\pi R}
\nonumber 
\end{equation}
is the boson linear density along the ring. The collapse rate \(\Gamma(n)\) is taken to activate only beyond a critical density \(n_c\), for example
\begin{equation}
\Gamma(n)
=
\Gamma_0\,
\Theta(n-n_c)
\left(
\frac{n}{n_c}-1
\right)^p,
\qquad p>0.
\nonumber 
\end{equation}
Motivated by the causality estimate derived from the projected exchange dynamics, a natural scale is
\begin{equation}
n_c=\frac{M c}{2\hbar}.
\nonumber 
\end{equation}
Equation (\ref{master_ring_collapse}) preserves particle number and rotational invariance, while selectively suppressing spatial coherences of the bosonic field.

The physical content of Eq.~(\ref{master_ring_collapse}) is transparent at the level of \(G^{(1)}\). Its collapse-induced contribution obeys
\begin{align}
\partial_t G^{(1)}(\theta,\theta';t)\big|_{\rm coll}
= 
-\Gamma(n)\,
\Bigl[
K_\ell(0)-K_\ell(\theta-\theta')
\Bigr]
\,G^{(1)}(\theta,\theta';t).
\end{align}
Hence the local density, corresponding to \(\theta=\theta'\), is unaffected, while the off-diagonal spatial coherence decays exponentially. The solution may be written as
\begin{equation}
G^{(1)}(\theta,\theta';t)
=
e^{-\Gamma(n)\,[K_\ell(0)-K_\ell(\theta-\theta')]\,t}
\,
G^{(1)}_{\rm unitary}(\theta,\theta';t),
\end{equation}
where \(G^{(1)}_{\rm unitary}\) denotes the coherence function in the absence of collapse.

For a ring geometry, a convenient periodic choice is
\begin{equation}
K_\ell(\Delta\theta)
=
\exp\!\left[
-\frac{2R^2\sin^2(\Delta\theta/2)}{\ell^2}
\right].
\end{equation}
This yields
\begin{align}
G^{(1)}(\theta,\theta';t) & =
\exp\left\{
-\Gamma(n)
\left[
1-
\exp\left(
-\frac{2R^2}{\ell^2}\sin^2\frac{\theta-\theta'}{2}
\right)
\right]t
\right\}
\times G^{(1)}_{\rm unitary}(\theta,\theta';t).
\label{G1_solution_kernel}
\end{align}
Equation~\eqref{G1_solution_kernel} shows explicitly that coherence between nearby points is weakly affected, whereas coherence between points separated by distances larger than \(\ell\) decays at a rate set by \(\Gamma(n)\).

This phenomenological modification has a direct implication for the projected spin dynamics discussed above. The coherent qubit exchange relies on the bosons acting as a single delocalized mediator around the entire ring. Once the density exceeds \(n_c\) and the coherence length falls below the circumference, \(\ell \ll 2\pi R\), the bosonic state no longer supports the global phase coherence required for this collective mediation. The apparent \(N\)-enhanced exchange is then dynamically suppressed, not by imposing a deterministic cutoff on the qubit dynamics, but by the loss of long-range first-order coherence in the bosonic ring itself.

We emphasize that Eq.~(\ref{master_ring_collapse}) should be regarded as a minimal nonrelativistic phenomenological model. Its purpose is not to provide a complete fundamental theory of collapse, but to encode in the simplest possible way the hypothesis that beyond a critical density the bosons cannot maintain global coherence over the ring, thereby invalidating the extrapolation of the fully coherent projected Hamiltonian into the large-\(N\) regime.

\section*{Discussion}

The proposed setup separates two ingredients that are often conflated in discussions of nonlocal electromagnetic effects: confinement of the magnetic field and nonlocal dependence of the superconducting phase on a linked fluxoid. In the toroidal geometry the magnetic field can remain localized inside the cores, while the loop energy depends on the total linked flux. This gives a concrete and experimentally accessible Hamiltonian in which two remote toroidal flux variables interact through a gauge-sensitive boundary condition.

The mode splitting in Eq.~\eqref{eq:normal_modes_identical}, the qubit coupling in Eq.~\eqref{eq:two_qubit_hamiltonian}, and their dependence on the linking orientation provide specific signatures. These observables should disappear when the common linking path is removed, and should be insensitive to smooth deformations that preserve the topology of the link. The primary experimental test is therefore twofold: correlated, time-resolved flux readout at the two toroids and a simultaneous null measurement of magnetic-field leakage in the region between them.

The apparent signaling paradox is not presented as a claim that controllable superluminal communication occurs. Rather, it is used as a diagnostic of the assumptions hidden in an instantaneous reduced Hamiltonian. The no-communication theorem assumes local operations generated by local Hamiltonians. A fluxoid model, after projection to a fixed global sector, violates that premise at the level of the effective description. The relevant physical question is where that description fails when embedded in a complete electrodynamic system.

The phenomenological master equation in Eq.~(\ref{master_ring_collapse}) is not proposed as a complete collapse theory. Its role is to parametrize an experimentally testable axiom: loss of long-range first-order coherence of the mediator can suppress the collective exchange before the reduced model reaches an apparent spacelike regime. This makes the architecture useful for testing fundamental collapse mechanisms. A null result would constrain the parameter space of density-dependent coherence-loss models; an observed suppression beyond known environmental channels would motivate a more complete microscopic theory.

The proposal therefore has two outcomes of interest. As a superconducting-circuit experiment, it realizes a fluxoid-mediated interaction between field-confined toroidal modes with orientation-tunable coupling. As a foundational probe, it converts questions about macroscopic spatiotemporal coherence into measurable circuit observables: mode splittings, time-resolved flux correlations, stray-field null tests, and density-dependent loss of coherent mediation.

\section*{Methods}

\subsection*{Response and normal modes}

For harmonic toroidal modes, Hamilton's equations or the Heisenberg equations \(\dot O=i[H,O]/\hbar\) give
\begin{equation}
\dot{\hat\phi}_j=\frac{\hat q_j}{C_j},
\qquad
\dot{\hat q}_2=-K_2\hat\phi_2-J\hat\phi_3+F_2,
\qquad
\dot{\hat q}_3=-K_3\hat\phi_3-J\hat\phi_2+F_3 .
\end{equation}
Combining these equations gives the second-order equations in the Results. Linearizing about the equilibrium point and Fourier transforming yields
\begin{equation}
\begin{pmatrix}
K_2-C_2\Omega^2 & J\\
J & K_3-C_3\Omega^2
\end{pmatrix}
\begin{pmatrix}
\delta\phi_2\\
\delta\phi_3
\end{pmatrix}
=0 .
\end{equation}
The normal modes therefore obey Eq.~\eqref{eq:normal_mode_equation}. For identical oscillators this reduces to Eq.~\eqref{eq:normal_modes_identical}. If oscillator 2 is driven externally and the backaction of oscillator 3 is ignored, the second row gives Eq.~\eqref{eq:transfer_function}.

\subsection*{Projection to two flux states}

For a toroidal element operated as a flux qubit, we project the flux coordinate to the two lowest persistent-current states,
\begin{equation}
\hat{\phi}_j\to \phi_j^\ast\sigma_j^z .
\end{equation}
The local tunnelling splitting is represented by \(-\Delta_j\sigma_j^x/2\). The self-terms \((\hat\phi_j)^2/(2L)\) become constants in the ideal two-state projection. The cross term becomes \(J_{zz}\sigma_2^z\sigma_3^z\), with \(J_{zz}=s_2s_3\phi_2^\ast\phi_3^\ast/L\). This gives Eq.~\eqref{eq:two_qubit_hamiltonian}.

\subsection*{Flux dynamics and Faraday fields}

It is important to distinguish a change of relative phase in the flux basis from a genuine change of the trapped flux. Let \(\hat\Phi_i=\Phi_q\hat n_i=\Phi_q(1-\sigma_z^{(i)})/2\) denote the flux trapped inside the \(i\)th toroidal element. A coherent evolution
\[
\ket{+}_i=\frac{\ket{0}_i+\ket{1}_i}{\sqrt2}
\longrightarrow
\ket{-}_i=\frac{\ket{0}_i-\ket{1}_i}{\sqrt2}
\]
does not, by itself, correspond to a time-dependent magnetic flux. If this evolution is generated by a Hamiltonian diagonal in the flux basis, then
\[
[\hat H_i,\hat\Phi_i]=0,
\qquad
\dot{\hat\Phi}_i=\frac{i}{\hbar}[\hat H_i,\hat\Phi_i]=0,
\]
and no Faraday electric field is induced. The common superconducting loop still couples to the flux through the static Aharonov--Bohm holonomy,
\[
\hat H_{\rm loop}^{(n)}
=
\frac{
\bigl(n\Phi_0-\Phi_b-\hat\Phi_2-\hat\Phi_3\bigr)^2
}{2L_{\rm loop}},
\]
but the sign change \(\ket{+}\to\ket{-}\) is a change of phase coherence, not a flux-transfer event.

By contrast, an evolution between flux eigenstates,
\[
\ket{0}_i\longrightarrow \ket{1}_i,
\]
requires a Hamiltonian component that does not commute with \(\hat\Phi_i\). In that case the trapped flux becomes a dynamical quantum operator and the induced electromotive force is
\[
\hat{\mathcal E}_i
=
-\frac{d\hat\Phi_i}{dt}
=
-\frac{i}{\hbar}[\hat H_i,\hat\Phi_i].
\]
For example, for a tunneling term \(\hat H_i=-(\Delta_i/2)\sigma_x^{(i)}\), one obtains
\[
\hat{\mathcal E}_i
=
-\frac{\Delta_i\Phi_q}{2\hbar}\sigma_y^{(i)} .
\]

When the toroidal flux changes between flux eigenstates, \(\ket{0}_i\rightarrow\ket{1}_i\), the process is different from the pure phase evolution \(\ket{+}_i\rightarrow\ket{-}_i\). Although the magnetic field remains confined inside the toroidal superconductor, the enclosed flux seen by the common loop changes in time. By Faraday's law this produces an azimuthal electric field on the common loop,
\begin{equation}
\oint_{\rm loop}\mathbf E\cdot d\mathbf l
=
-\frac{d}{dt}
\left(
\Phi_b+\hat\Phi_2+\hat\Phi_3
\right).
\end{equation}
For a circular loop of radius \(R\), in the gauge where the AB vector potential is distributed uniformly along the loop,
\begin{equation}
A_\varphi(t)
=
\frac{
\Phi_b+\hat\Phi_2(t)+\hat\Phi_3(t)
}{2\pi R},
\qquad
E_\varphi(t)
=
-\partial_t A_\varphi(t)
=
-\frac{
\dot{\hat\Phi}_2(t)+\dot{\hat\Phi}_3(t)
}{2\pi R}.
\end{equation}
Thus a \(0\rightarrow1\) flux transition produces an electric-field pulse whose time integral is fixed by the change of trapped flux,
\begin{equation}
\int dt\,E_\varphi(t)
=
-\frac{\Delta\Phi_i}{2\pi R}.
\end{equation}
This field does not arise from magnetic-field leakage out of the toroid; it is the Faraday field associated with the time-dependent AB holonomy.

Consequently, the loop Hamiltonian is not modified by adding an independent capacitive degree of freedom, but by making the AB flux entering the fluxoid energy explicitly time dependent. In a fixed fluxoid sector \(n\), the common superconducting loop is described by
\begin{equation}
\hat H_{\rm loop}^{(n)}(t)
=
\frac{
\left[
n\Phi_0-\Phi_b-\hat\Phi_2(t)-\hat\Phi_3(t)
\right]^2
}{2L_{\rm loop}} .
\end{equation}
Equivalently, for a charged-particle ring representation, the same Faraday effect appears through the time-dependent minimal-coupling Hamiltonian
\begin{equation}
\hat H_{\rm ring}(t)
=
E_R
\sum_{m\in\mathbb Z}
\left[
m-
\frac{
\Phi_b+\hat\Phi_2(t)+\hat\Phi_3(t)
}{\Phi_0}
\right]^2
\hat b_m^\dagger \hat b_m .
\end{equation}
The electric field therefore enters through the time dependence of the Aharonov--Bohm flux, or equivalently through the time-dependent vector potential around the loop.

In contrast, a coherent transformation \(\ket{+}_i\rightarrow\ket{-}_i\) generated by a Hamiltonian diagonal in the flux basis has
\begin{equation}
\dot{\hat\Phi}_i=0,
\end{equation}
and therefore produces no Faraday field on the common loop. It changes only the relative phase between the two flux components. A genuine electric-field pulse is produced only when the flux eigenvalue itself changes, as in a \(\ket{0}_i\rightarrow\ket{1}_i\) process.

\subsection*{Covariant form of the causal coherence bound}
\label{sec:covariant_collapse}

The collapse criterion introduced above was written in the rest frame of the
device. The collapse or the nonlinear evolution should be frame independent. We now give a covariant formulation whose nonrelativistic limit
reduces to that device-frame master equation. We use the metric convention
\(g_{\mu\nu}=(-,+,+,+)\). The superconducting condensate is described by a
gauge-invariant Cooper-pair four-current
\begin{equation}
J^\mu(x)=q_\ast n_0(x)u^\mu(x),
\label{eq:Jmu_def}
\end{equation}
where \(q_\ast=2e\), \(u^\mu\) is the local condensate four-velocity, and
\(n_0\) is the proper condensate density. Current conservation gives
\begin{equation}
\nabla_\mu J^\mu=0 .
\end{equation}
The proper density is the Lorentz scalar
\begin{equation}
n_0(x)
=
\frac{1}{q_\ast c}
\sqrt{-J^\mu(x)J_\mu(x)} ,
\label{eq:n0_scalar}
\end{equation}
and the condensate four-velocity is
\begin{equation}
u^\mu(x)
=
\frac{J^\mu(x)}{q_\ast n_0(x)} ,
\qquad
u^\mu u_\mu=-c^2 .
\label{eq:u_def}
\end{equation}
The tensor
\begin{equation}
h_{\mu\nu}
=
g_{\mu\nu}
+
\frac{u_\mu u_\nu}{c^2}
\label{eq:spatial_projector}
\end{equation}
projects onto directions locally orthogonal to the condensate flow and
therefore defines proper spatial distances in the condensate rest frame.

For a coherent mediator, the relevant density entering the
one-dimensional causal bound is the proper linear condensate density along the
link. If \(\mathcal A_0\) denotes the cross section measured in the local rest
frame of the condensate, we define
\begin{equation}
\lambda_0
=
\int_{\mathcal A_0} dA_0\, n_0 .
\label{eq:lambda0_def}
\end{equation}
For a uniform wire this reduces to \(\lambda_0=n_0A_0\). The proper coherent
length \(L_0\) is the spatial extent of the coherent condensate measured with
the projector \(h_{\mu\nu}\), while \(\tau_0\) is the proper duration over
which the same many-body condensate state remains phase coherent. In this
language the device-frame collapse parameter is promoted to the scalar
\begin{equation}
\chi_{\rm cov}
=
\frac{\lambda_0}{\lambda_c}
\frac{c\tau_0}{L_0},
\label{eq:chi_cov}
\end{equation}
where the critical proper linear density is
\begin{equation}
\lambda_c
=
\frac{M_\ast c}{h}.
\label{eq:lambda_c}
\end{equation}
Equation~\eqref{eq:chi_cov}
is invariant because \(\lambda_0\), \(\lambda_c\), \(L_0\), and \(\tau_0\)
are all defined as proper quantities in the local condensate rest frame. In
the laboratory frame of a stationary device,
\begin{equation}
\lambda_0\rightarrow n_{1D},\qquad
L_0\rightarrow L_{\rm coh},\qquad
\tau_0\rightarrow \tau_{\rm coh},
\end{equation}
so Eq.~\eqref{eq:chi_cov} reduces to the nonrelativistic parameter
\begin{equation}
\chi_{\rm caus}
=
\frac{n_{1D}}{n_{1D}^{\rm crit}}
\frac{c\tau_{\rm coh}}{L_{\rm coh}} .
\end{equation}

The collapse operator should not be the ordinary mass density, as in GRW-like
or standard CSL-like models, because the relevant degree of freedom in the
present proposal is not a spatial separation of massive matter. It is the
long-range gauge coherence of a superconducting condensate enforcing a fluxoid
constraint. We therefore choose collapse operators constructed from
gauge-invariant condensate current and fluxoid variables. Locally, the
rest-frame condensate density operator may be written as the scalar
\begin{equation}
\hat{\mathcal N}(x)
=
-\frac{u_\mu(x)\hat J^\mu(x)}{q_\ast c^2},
\label{eq:N_operator_cov}
\end{equation}
which reduces to the Cooper-pair density operator in the condensate rest
frame. The associated gauge-invariant fluxoid phase around a closed
superconducting contour \(\mathcal C\) is
\begin{equation}
\hat{\mathcal F}_{\mathcal C}
=
\oint_{\mathcal C}
\left[
\nabla\hat\theta
-
\frac{q_\ast}{\hbar}\hat{\mathbf A}
\right]\cdot d\mathbf l ,
\label{eq:fluxoid_operator}
\end{equation}
or, equivalently,
\begin{equation}
\hat{\mathcal W}_{\mathcal C}
=
\exp\left(i\hat{\mathcal F}_{\mathcal C}\right).
\label{eq:Wilson_fluxoid}
\end{equation}
Both \(\hat{\mathcal F}_{\mathcal C}\) and
\(\hat{\mathcal W}_{\mathcal C}\) are invariant under the gauge
transformation
\begin{equation}
\hat\theta\rightarrow \hat\theta+\frac{q_\ast}{\hbar}\Lambda,
\qquad
\hat A_\mu\rightarrow \hat A_\mu+\partial_\mu\Lambda .
\end{equation}

A covariant phenomenological collapse equation can then be written on a
spacelike hypersurface \(\Sigma\). Let \(\rho_\Sigma\) be the density matrix
associated with \(\Sigma\). We take the collapse part of the Tomonaga--
Schwinger evolution to be
\begin{align}
\frac{\delta \rho_\Sigma}{\delta \Sigma(x)}
&=
-\frac{i}{\hbar}
[\hat{\mathcal H}(x),\rho_\Sigma]
\nonumber\\
&\quad
-\frac{\gamma_0}{2}
F\!\left[\chi_{\rm cov}(x)\right]
\int_{\Sigma} d\Sigma_y\,
K_\ell(x,y)
\left[
\hat{\mathcal N}(x),
\left[
\hat{\mathcal N}(y),
\rho_\Sigma
\right]
\right].
\label{eq:covariant_master}
\end{align}
Here \(K_\ell(x,y)\) is a positive smearing kernel of finite invariant width
\(\ell\), and \(F(\chi)\) is a threshold function satisfying
\begin{equation}
F(\chi)\simeq0 \quad {\rm for}\quad \chi<1,
\qquad
F(\chi)>0 \quad {\rm for}\quad \chi>1 .
\label{eq:F_chi}
\end{equation}
For example,
\begin{equation}
F(\chi)=\Theta(\chi-1)(\chi-1)^p,
\qquad p>0 .
\label{eq:F_example}
\end{equation}
The finite kernel \(K_\ell(x,y)\) prevents the collapse noise from being
strictly point-local and encodes the finite spacetime resolution of the
coherence-loss process. A convenient rest-frame form is
\begin{equation}
K_\ell(x,y)
=
\exp\left[-\frac{d_\perp^2(x,y)}{2\ell^2}\right],
\label{eq:K_cov}
\end{equation}
where
\begin{equation}
d_\perp^2(x,y)
=
h_{\mu\nu}(x-y)^\mu(x-y)^\nu
\label{eq:dperp}
\end{equation}
is the squared proper spatial distance projected orthogonally to the local
condensate four-velocity.

We now show that Eq.~\eqref{eq:covariant_master} reproduces the
nonrelativistic master equation used above. In the rest frame of a stationary
superconducting ring,
\begin{equation}
u^\mu=(c,\mathbf 0),
\qquad
\Sigma:t={\rm const.},
\end{equation}
and the scalar density operator reduces to
\begin{equation}
\hat{\mathcal N}(x)\rightarrow \hat n(\theta)
=
\hat\psi^\dagger(\theta)\hat\psi(\theta),
\end{equation}
where
\begin{equation}
\hat\psi(\theta)
=
\frac{1}{\sqrt{2\pi}}
\sum_{m\in\mathbb Z}
e^{im\theta}\hat b_m .
\end{equation}
For a ring of radius \(R\), the proper chord distance between two angular
positions is
\begin{equation}
d_\perp^2(\theta,\theta')
=
4R^2\sin^2\left(\frac{\theta-\theta'}{2}\right).
\end{equation}
Thus Eq.~\eqref{eq:K_cov} becomes
\begin{equation}
K_\ell(\theta-\theta')
=
\exp\left[
-\frac{2R^2}{\ell^2}
\sin^2\left(\frac{\theta-\theta'}{2}\right)
\right].
\label{eq:K_ring_reduction}
\end{equation}
Defining
\begin{equation}
\Gamma(n_{1D})
=
\gamma_0 F(\chi_{\rm caus}),
\end{equation}
Eq.~\eqref{eq:covariant_master} reduces to
\begin{align}
\partial_t\rho
&=
-\frac{i}{\hbar}[H,\rho]
\nonumber\\
&\quad
-\frac{\Gamma(n_{1D})}{2}
\int_0^{2\pi}d\theta
\int_0^{2\pi}d\theta'\,
K_\ell(\theta-\theta')
\left[
\hat n(\theta),
\left[
\hat n(\theta'),
\rho
\right]
\right].
\label{eq:nonrel_master_recovered}
\end{align}
This is precisely the density-dephasing master equation introduced in the
device-frame treatment.

Finally, the action of Eq.~\eqref{eq:nonrel_master_recovered} on the one-body
density matrix
\begin{equation}
G^{(1)}(\theta,\theta';t)
=
\langle
\hat\psi^\dagger(\theta,t)\hat\psi(\theta',t)
\rangle
\end{equation}
is obtained using
\begin{equation}
[\hat n(\alpha),\hat\psi(\theta)]
=
-\delta(\alpha-\theta)\hat\psi(\theta),
\qquad
[\hat n(\alpha),\hat\psi^\dagger(\theta)]
=
\delta(\alpha-\theta)\hat\psi^\dagger(\theta).
\end{equation}
The collapse contribution is
\begin{equation}
\partial_tG^{(1)}(\theta,\theta';t)\big|_{\rm coll}
=
-\Gamma(n_{1D})
\left[
K_\ell(0)-K_\ell(\theta-\theta')
\right]
G^{(1)}(\theta,\theta';t).
\label{eq:G1_decay_cov_reduction}
\end{equation}
Therefore
\begin{align}
G^{(1)}(\theta,\theta';t)
&=
\exp\left\{
-\Gamma(n_{1D})
\left[
1-
\exp\left(
-\frac{2R^2}{\ell^2}
\sin^2\frac{\theta-\theta'}{2}
\right)
\right]t
\right\}
\nonumber\\
&\quad\times
G^{(1)}_{\rm unitary}(\theta,\theta';t).
\label{eq:G1_solution_cov_reduction}
\end{align}
The local density, corresponding to \(\theta=\theta'\), is unchanged, whereas
off-diagonal long-range coherence is suppressed. Thus the covariant current-
based collapse law reduces in the device rest frame to the proposed
phenomenological loss of long-range condensate coherence. The model is
therefore distinct from mass-density localization models: its collapse
operator is tied to gauge-invariant condensate current and fluxoid coherence,
and its rate is controlled by the scalar spacetime-coherence parameter
\(\chi_{\rm cov}\).

\section*{Acknowledgements}

This work has been supported in part by the Laboratory Directed Research and Development program and Sandia University Partnerships Network ( SUPN) program and performed in part at the Center for Integrated Nanotechnologies ( CINT), an Office of Science User Facility operated for the U.S. Department of Energy (DOE) Office of Science. Sandia National Laboratories is a multimission laboratory managed and operated by National Technology and Engineering Solutions of Sandia, LLC., a wholly owned subsidiary of Honeywell International, Inc., for the U.S. Department of Energy's National Nuclear Security Administration under Contract No. DE-NA0003525. This article describes objective technical results and analysis. Any subjective views or opinions that might be expressed in the article do not necessarily represent the views of the U.S. Department of Energy or the United States Government.

\section*{Competing interests}

The authors declare no financial or non-financial competing interests.

\bibliographystyle{unsrtnat}
\bibliography{References,references_npj}



\end{document}